# Analyzing the Dynamics of COVID-19 Lockdown Success: Insights from Regional Data and Public Health Measures

Md. Motaleb Hossen Manik, Md. Ahsan Habib, Md. Zabirul Islam, Tanim Ahmed, and Fabliha Haque
*Department of Computer Science and Engineering (CSE)*
*Khulna University of Engineering & Technology*
Khulna-9203, Bangladesh
mh.manik@cse.kuet.ac.bd, mahabib@cse.kuet.ac.bd, zabir.kuetcse@gmail.com, tanimahmedcse@gmail.com, and fablihahaque.kuet@gmail.com

*Abstract*— The COVID-19 pandemic caused by the coronavirus had a significant effect on social, economic, and health systems globally. The virus emerged in Wuhan, China, and spread worldwide resulting in severe disease, death, and social interference. Countries implemented lockdowns in various regions to limit the spread of the virus. Some of them were successful and some failed. Here, several factors played a vital role in their success. But mostly these factors and their correlations remained unidentified. In this paper, we unlocked those factors that contributed to the success of lockdown during the COVID-19 pandemic and explored the correlations among them. Moreover, this paper proposes several strategies to control any pandemic situation in the future. Here, it explores the relationships among variables, such as population density, number of infected, death, recovered patients, and the success or failure of the lockdown in different regions of the world. The findings suggest a strong correlation among these factors and indicate that the spread of similar kinds of viruses can be reduced in the future by implementing several safety measures.

## I. INTRODUCTION

The SARS-CoV-2 virus that caused the COVID-19 pandemic had a big impact on the global economy, social activities, and health sectors. It was identified in late 2019 in Wuhan, China, and since then, it has become a global pandemic [1]. Since its first identification, it spread rapidly across countries. Moreover, COVID-19 is highly contagious and leads to severe illness or death, particularly in elderly individuals or those with pre-existing health conditions [2]. It spreads through respiratory droplets (coughs or sneezes), close contact with an infected person, touching infected surfaces, etc. [3]. Almost all governments throughout the world implemented travel restrictions, lockdowns, and mask policies in response to the pandemic to stop the virus from spreading [4], [5]. However, the pandemic exposed existing inconsistencies and inequalities in healthcare and other areas. The healthcare systems faced unprecedented challenges, with shortages of medical supplies and overwhelmed hospitals in some regions [6], [7], [8].

Several measures were taken worldwide to limit the spread of COVID-19, including washing hands frequently with soap, covering mouth and nose while coughing or sneezing, avoiding close contact with sick people, staying at home, etc. [9]. Despite enforcing these measures, many countries could not completely stop the spread of the virus [10]. The number of infected, dead, and recovered patients increased daily basis. Multiple factors played a vital role in this growth [11]. Moreover, these factors were correlated with each other where some had a higher impact on the spread of the virus and some had a lower. Among them, the population density of any region was a noticeable one [12]. Moreover, some lockdowns were a success and some were a failure. Here, several factors played a vital role in the success of the lockdowns. Many ways were introduced to stop the pandemic [13], [14], [15], [16]

One significant finding is higher population density which can increase the risk of any virus transmission because viruses spread more easily when people are close to each other [12]. Furthermore, implementing social distancing measures and providing proper access to testing, personal protective equipment (PPE), and healthcare services are more difficult in densely populated regions. Additionally, population density is not the only factor that contributes to the transmission of any virus. Other factors, such as the efficiency of public health interventions, community involvement in preventing viral spread, and the prevalence of underlying health issues in the population can also play a role in virus transmission.

One major objective of this study is to investigate the correlation among different factors that affected the lockdown measures implemented during the COVID-19 pandemic. The study aims to understand the relationship among variables such as population density, number of infected, deaths, and recoveries, and their impact on the effectiveness of lockdowns in various regions of the world. Through an analysis of these factors, the study proposes future strategies for controlling and managing similar pandemic situations. The findings of the research highlight the importance of maintaining precautionary measures to reduce the spread of similar kinds of viruses in the future.

The study makes the following major contributions.

*i)* To develop a comprehensive dataset for COVID-19-related research by collecting data considering multiple attributes from different regions of the world.

*ii)* To explore the factors that were implemented worldwide and contributed to the success of lockdown measures.

*iii)* To investigate the correlations between those factors and analyze them mathematically.

*iv)* Finally, to propose a protective direction to mitigate and manage any future pandemic situations.

The rest of the paper is organized as follows. Section II reviews the related works. Section III describes the materials and methodology. Section IV represents the results and findings. Finally, section V concludes the paper with some future directions.

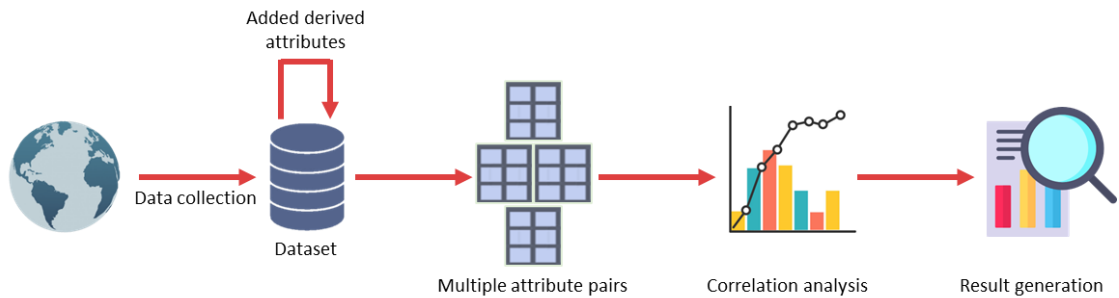

Fig. 1: Overall architecture of the proposed framework.

## II. RELATED WORKS

Many research articles have been published from multiple perspectives including examining measures taken during COVID-19, determining the correlation between the factors behind the success or failure of lockdowns during the pandemic [17], etc. Some studies have explored the precautionary measures behind the success or failure of lockdowns [18], [19], [20]. In contrast, others have attempted to extract the correlations between those factors [11], [12].

Dawoud D. [21] has studied the variety of measures taken by different countries to avoid the second wave of COVID-19. The author has shown that the primary aim of the different governments was to flatten the pandemic curve during the pandemic situation rather than initially taking any action against coronavirus. This was conducted to enhance their healthcare systems for the enormous demand that has never been seen in peacetime. Moreover, the author has suggested some measures that can facilitate the loosening of the spread of the virus. These include increasing the use of antigen testing, introducing antibody testing, using innovative technologies widely, maintaining social distance in public gatherings, covering one's face in public places, providing frontline workers with adequate PPE, providing access to the COVID-19 vaccination program, developing public awareness, etc. However, this work only studied some measures to limit the spread of the virus rather than finding any correlation between different factors that played a vital role in the success against coronavirus.

Another study on the success of lockdown in Greece was presented in [20]. The authors studied the correlation between multiple factors and lockdown success after the lockdown was implemented in Greek society. They performed linear regression on Delta Day (DD), confirmed cases, and the number of reported mortalities. Their findings suggested that higher healthcare costs as a percentage of the national GDP were not associated with greater 30-day mortality outcomes. Again, they discovered that the DD index was substantially linked to COVID-19 incidence per million persons at 30 days (p =0.001). The link between DD and the 30-day death rate per million persons was not statistically significant (p =0.087), indicating that other variables contributed to COVID-19-related mortality.

A similar study was conducted in [19], in which the authors studied the challenges and success of the COVID-19 lockdown in South Africa. They presented the impact of the lockdown and how the South African government tackled its outcomes. Although the number of infected patients was kept at a limit, the situation in society became unsteady. Problems such as food supply, riots, medicine scarcity, job loss, financial issues, and lack of proper sanitation were all over society. Still, they succeeded in flattening the curve of the pandemic and ending the lockdown in 'phase 5.' The government provided food and reduced lockdowns at a low rate. They transformed the stadiums into hospitals and provided small and medium-sized businesses with financial support. Although this study showed the method of lockdown success, it did not focus on the correlation between the measures that accelerated success.

Cole et al. [22] provided a system with two steps-based models, where the first phase employs machine learning and the second phase uses the Augmented Synthetic Control Model, to assess the effects of lockdown. Their detailed approach demonstrated the impact of lockdowns on public health and air pollution. They calculated that reducing $NO_2$ levels may have prevented 10,822 deaths in China. But their model left out any further elements that could have contributed to the lockdown's success throughout the outbreak.

Another type of work on determining the impact of lockdown on different aged people has been proposed in [23], [24], [25]. Authors of [23] carried out their study on 389 young people and determined that the majority of them suffered from depression, which was 55% greater than before the epidemic. Again, the authors of [24] carried out their research on 100 respondents, focusing on six difficulties: family, sleep disorders, future anxiety, anger, a lack of emotional support, and dread of receiving bad news. The authors discovered that for these six criteria, the participants experienced 90%, 86%, 85%, 83%, 79%, and 72% stress and anxiety, respectively. The authors of [25] concentrated on students' social networks and mental health throughout the epidemic. They concluded that the lockdown had harmed about 39.2 percent of the pupils. This research examined a variety of individuals' reactions to the lockdown, but they failed to show a connection between their reactions and the variables that made the lockdown successful or unsuccessful.

## III. MATERIALS AND METHODS

The proposed architecture is thoroughly described in this section. The architecture of the proposed framework is illustrated in fig. 1. Initially, a dataset was created by collecting data regarding different aspects (number of infected, death and recovered patients, area, population, etc.) of various countries, and later the successful existence of lockdowns in certain locations of the countries was included in the dataset. Since the correlation coefficient is an established method for evaluating relationships among different aspects, we employed a variety of correlation approaches (Pearson's correlation, Spearman's rho, and Kendall's tau) to analyze the attributes of the dataset and determine their intercorrelations.

TABLE I. SAMPLE DATASET

| Region Name | Infected | Deaths | Recovered | Area | Population | Success/Failure |
|---|---|---|---|---|---|---|
| Alaska | 761 | 12 | 491 | 665384 | 731545 | success |
| Iowa | 28819 | 709 | 22870 | 56272 | 3155070 | failure |
| New York | 393454 | 24855 | 70590 | 54554 | 19453561 | failure |
| Maharashtra | 284281 | 11194 | 158140 | 307713 | 112374333 | success |
| North Dakota | 3313 | 77 | 2952 | 70698 | 762062 | success |

TABLE II. DERIVED DATASET

| Region Name | Infected | Death | Recovered | Area | Population | Success/Failure | Density |
|---|---|---|---|---|---|---|---|
| Alaska | 761 | 12 | 491 | 665384 | 731545 | 1 | 1.0994 |
| Iowa | 28819 | 709 | 22870 | 56272 | 3155070 | 0 | 56.0682 |
| New York | 393454 | 24855 | 70590 | 54554 | 19453561 | 0 | 356.5927 |
| Maharashtra | 284281 | 11194 | 158140 | 307713 | 112374333 | 1 | 356.1920 |
| North Dakota | 3313 | 77 | 2952 | 70698 | 762062 | 1 | 10.7791 |

## A. Data Collection

Since the desired dataset is new in terms of the objective of this research, the dataset was manually generated by collecting data from this source [26] that contains relevant data that aligned with the research. The created dataset consists of seven features: *i) the name of the region*, *ii) the number of infected*, *iii) the number of death* and *iv) the number of recovered patients in that region*, *v) the area of that region*, *vi) the population*, and *vii) the successful launch of lockdown on that region*. The collection includes 10,000 data entries up to December 2022 that were collected from 100 different countries. The sample dataset with seven features is presented in Table I. The raw data has also been verified repeatedly to eliminate duplicates and inconsistencies. The success and failure (last attribute) have been chosen based on a study [18] of the lockdown's success rate in a particular region.

## B. Dataset Preprocessing

Here, data normalization has been carried out for each numeric attribute using the min-max (*min* = 0 and *max* = 1) normalization method to maintain the values within the range of 0 and 1. Since the information was manually collected by the authors and did not contain the density of the regions, the density attribute has been derived from the population and area of each region. Since the success or the failure is not numeric values, the successes have been replaced by 1 and the failures have been replaced by 0. To eliminate biases, the dataset has been generated with an equal distribution of samples (5000 *successes* and 5000 *failures*). After adding the derived attribute, Table II shows the dataset's actual version.

## C. Correlation Coefficient

A correlation coefficient summarizes whether a change in one variable is associated with a change in other variables. This statistic is descriptive. Therefore, it presents a summary of a sample of data without drawing any conclusions about the complete population. When it describes the link between two variables, a correlation coefficient is considered a bivariate statistic. However, when it does so among more than two variables, it is considered a multivariate statistic.

When a variable increases with the increase of another variable, they are said to be positively correlated, resulting in a positive correlation coefficient. Conversely, if one variable decreases as the other increases, it is referred to as a negative correlation, yielding a negative correlation coefficient. If the variables are independent and have no relationship, a correlation coefficient of zero is generated.

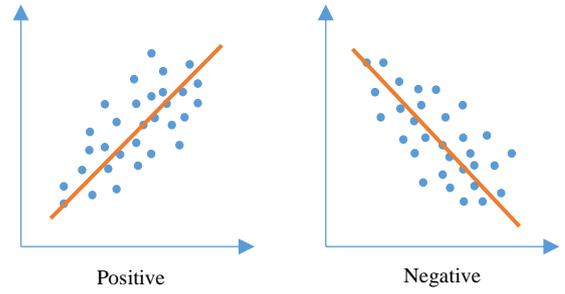

Fig. 2: Types of correlation.

To determine the linear or non-linear relationship between variables, a visually plotted dataset can be employed. While a linear pattern denotes that a straight line is the best match between the data points, a non-linear or curvilinear pattern can take any number of other shapes, including a U-shape or a line with a curve. Fig. 2 depicts an example of positively and negatively correlated variables with linear fitted curves.

In this study, various correlation coefficient approaches have been utilized to determine the relationship among the attributes that are mentioned in Sec. III(A).

## D. Categories of Correlation Coefficient

The correlation coefficient is categorized into various types based on the data distribution, the linearity of the relationships among the data, and the measurement level of the variables. This paper utilizes several of these types, and their details are as follows:

*i)* *Pearson's r*: It is usually referred to as *Pearson's product-moment correlation coefficient*, indicating the linear relationship between two quantitative variables. It is calculated by the following equation.

$$r = \frac{n \sum xy - (\sum x)(\sum y)}{\sqrt{[n \sum x^2 - (\sum x)^2][n \sum y^2 - (\sum y)^2]}} \quad (1)$$

Here,

- $r$ = correlation coefficient between x and y
- $x$ and $y$ = values of *X* and *Y* variables
- $n$ = sample size

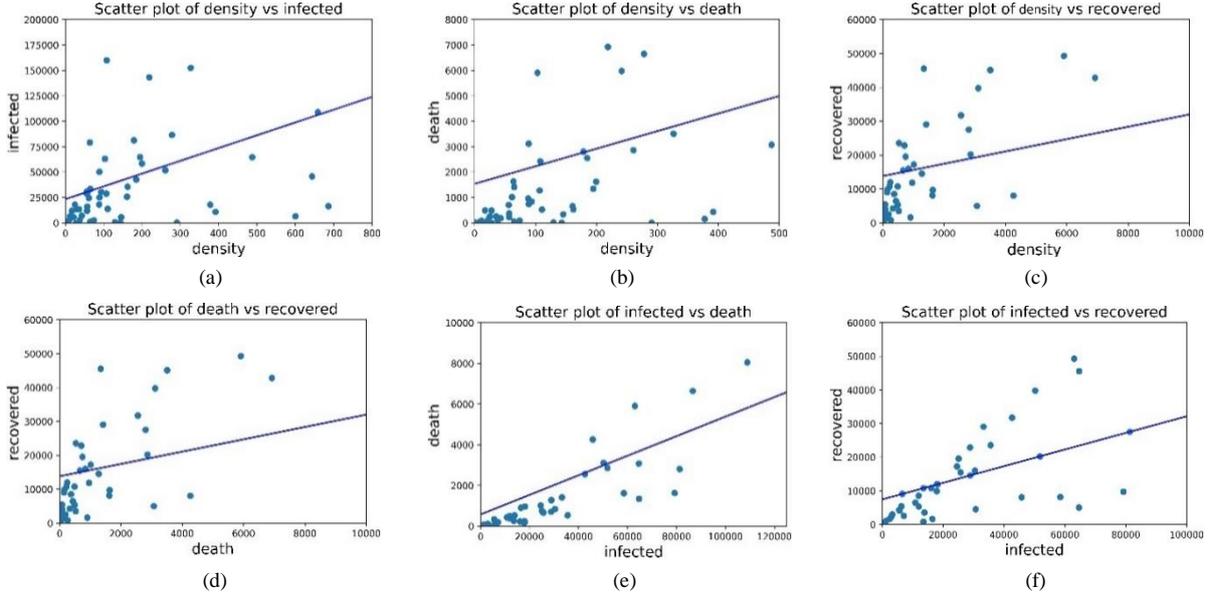

Fig. 3: Scatter plot of *(a) Density vs Infected, (b) Density vs Death, (c) Density vs Recovered, (d) Death vs Recovered, (e) Infected vs Death, and (f) Infected vs Recovered*.

*ii) Spearman's Rho:* It is the most preferred alternative for *Pearson's r* coefficient. Rather than using the raw data directly for calculation, the rank correlation coefficient employs the rank of the data from each variable (*e.g.*, from lowest to highest). While the *Pearson correlation coefficient* measures associations' linearity, the Spearman correlation coefficient evaluates the monotonicity of correlations. Each variable moves in one direction at the same pace over the whole data range when there is a linear relationship. However, every variable in a monotonic connection changes, but not always at the same pace. It is calculated by the following equation.

$$r = 1 - \frac{6 \sum d_i^2}{\sqrt{(n^3 - n)}} \qquad (2)$$

Here,

- $r$ = strength of the rank correlation between variables
- $d_i$ = the difference between the *X*-variable rank and the *Y*-variable rank for each pair of data
- $n$ = sample size

To use this formula, the data is ranked from each variable separately from low value to high value. Then, the differences ($d_i$) between the ranks of the variables for each data pair are calculated and provided as the main input for the above formula.

*iii) Kendall's Tau:* Kendall's rank correlation (*τ*) is a distribution-free test of independence and an assessment of the degree of dependence between two variables. Even if Spearman's rank correlation is sufficient for examining the null hypothesis of independence between two variables, it can occasionally be challenging to understand why the null hypothesis was found to be incorrect. By highlighting the level of reliance between the variables under examination, Kendall's rank correlation improves on this. The equation below is used to compute it.

$$\tau = \frac{n_c - n_d}{n_0} \qquad (3)$$

Here,

- $n_c$ = number of concordant pairs
- $n_d$ = number of discordant pairs
- $n_0$ = total number of pairs

IV. RESULTS AND FINDINGS

This section describes the experimental measures, setup, results, and findings of the proposed framework. These sub-sections are as follows:

*A. Experimental Measurements*

This experiment has been carried out on the dataset discussed in Sec. III (A). The dataset contains eight significant attributes. From these attributes, six different combinations have been considered as the experimental measurements. The combinations are as follows:

i) (*Density, Infected*)

ii) (*Density, Death*)

iii) (*Density, Recovered*)

iv) (*Death, Recovered*)

v) (*Infected, Death*)

vi) (*Infected, Recovered*)

*B. Experimental Setup*

This experiment was conducted using Python programming language on the Kaggle online platform. It was carried out on a desktop computer containing Windows 11 with an Intel Core i5 processor running at 3.10 GHz and 12 GB of RAM. Though several complex mathematical operations were needed in the model, still no parallel processing was applied to the experiment.

## C. Experimental Results

This section has been divided into five parts, *i.e.*, *i) plotting the data points*, *ii) Pearson's correlation coefficient*, *iii) Spearman's correlation coefficient*, *iv) Kendall's correlation coefficient,* and *v) success of lockdown*. They are described below.

*i) Plotting the data points*: Initially, the data points are illustrated with a scatter plot for the six combinations of attributes as mentioned in Sec. IV (A). Fig. 3 represents the corresponding scatter plots. Here, in each combination, it is visible that the corresponding attributes are positively correlated. However, determining the strength of correlation only by examining the scatter plot is difficult. Here, the correlation coefficient helps to find out the strength of the correlations.

*ii) Pearson's correlation coefficient*: The value of *r* for Pearson's correlation according to (1) has been calculated for the six combinations of attributes. Table III shows the corresponding values of those combinations. The remarks about the strength of correlation [27] for each combination have also been mentioned as an indication.

TABLE III: PEARSON'S CORRELATION COEFFICIENT FOR DIFFERENT COMBINATIONS OF ATTRIBUTES

| Combination | Value of *r* | Strength of correlation |
|---|---|---|
| *Density vs Infected* | 0.384 | Moderate |
| *Density vs Death* | 0.284 | Weak |
| *Density vs Recovered* | 0.277 | Weak |
| *Death vs Recovered* | 0.412 | Moderate |
| *Infected vs Death* | 0.649 | Moderate |
| *Infected vs Recovered* | 0.755 | Strong |

*iii) Spearman's correlation coefficient:* The value of *r* for Spearman's correlation has been calculated using (2) for the six combinations and table IV illustrates the values. The remarks about the strength of correlation [27] for each combination have also been mentioned here as an indication.

TABLE IV: SPEARMAN'S CORRELATION COEFFICIENT FOR DIFFERENT COMBINATIONS OF ATTRIBUTES

| Combination | Value of *r* | Strength of correlation |
|---|---|---|
| *Density vs Infected* | 0.54 | Moderate |
| *Density vs Death* | 0.44 | Moderate |
| *Density vs Recovered* | 0.42 | Moderate |
| *Death vs Recovered* | 0.724 | Strong |
| *Infected vs Death* | 0.864 | Strong |
| *Infected vs Recovered* | 0.877 | Strong |

*iv) Kendall's correlation coefficient:* The value of τ for Kendall's correlation has been calculated for the six combinations using (3) and table V presents the corresponding values. Here, again the remarks about the strength of correlation [27] for each combination have been mentioned here as an indication.

TABLE V: KENDALL'S CORRELATION COEFFICIENT FOR DIFFERENT COMBINATIONS OF ATTRIBUTES

| Combination | Value of *r* | Strength of correlation |
|---|---|---|
| *Density vs Infected* | 0.413 | Moderate |
| *Density vs Death* | 0.347 | Weak |
| *Density vs Recovered* | 0.318 | Weak |
| *Death vs Recovered* | 0.582 | Moderate |
| *Infected vs Death* | 0.744 | Strong |
| *Infected vs Recovered* | 0.737 | Strong |

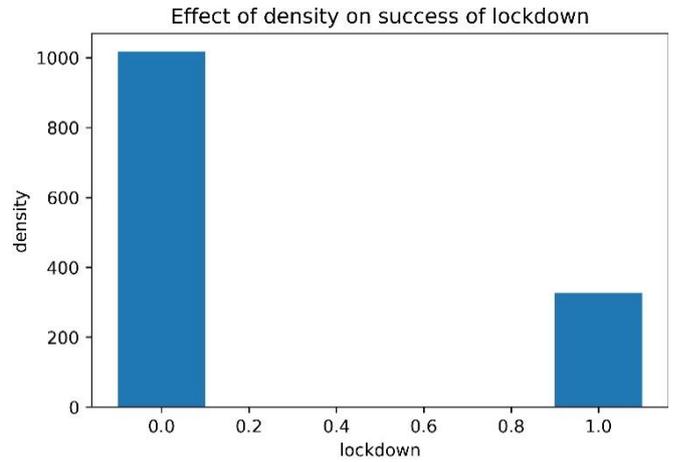

Fig. 4: Effect of density on success of lockdown.

*v) Success of lockdown:* Since different counties were enforcing lockdowns in different parts, some were successful (1) and some were a failure (0). Here, a visual representation is shown in fig. 4 which indicates that the success of lockdown is mostly seen in the regions with a lower density of population. Still, some regions with higher density tackled the coronavirus with different precautionary measures including proper supply of daily goods, and medicines, maintaining proper rules, complete separation of social engagement, etc. [28].

## D. Findings and directions

From Tables 3, 4, and 5, it can be observed that the combinations "infected vs death" and "infected vs recovered" have strong positive correlations. This indicates that in any region, the number of deaths and recovered patients increased with the number of infected people. Moreover, other combinations exhibited moderate and weak correlations. Among them, the combinations paired with density have weak correlations, which indicates that density had less effect on infected, dead, and recovered patients compared to other combinations. However, it cannot be concluded that density does not play a vital role in the number of infected, dead, and recovered patients. Although it has some impact, still it is moderate and weak in most cases [20]. Furthermore, fig. 4 demonstrates a strong correlation between the success [18] of the lockdown and the population density of the regions, which in turn is closely linked to the rates of infection, death, and recovery among patients. Hence, in the future, the spread of any pandemic situation can be controlled by reducing the number of infected and dead patients as well as comparatively increasing the number of recovered patients. Furthermore, a critical factor in reducing the spread of a pandemic is the reduction of population density in any given location. Decreasing population density can significantly contribute to mitigating the transmission of infectious diseases and limiting the rapid spread among inhabitants. By implementing measures to reduce population density, such as promoting remote work, implementing social distancing guidelines, and encouraging decentralized living arrangements, the future spread of pandemics can be effectively reduced and proper safety can be ensured.

## V. CONCLUSIONS

This research studies COVID-19 data from December 2019 to December 2022 and examines the success or failure of lockdowns enforced in different regions worldwide. Initially, the authors collected data by considering different factors related to COVID-19 and generated a comprehensive dataset that included the success or failure of lockdowns in various regions of the world. Then, different correlation-finding methods have been applied to the comprehensive dataset. This study shows that the number of deaths and recovered patients increased at a high rate with the number of infected patients. Moreover, the population density loosely affected the number of recovered patients. Again, the measures behind the success of the lockdown are mostly related to the population density of regions. So, by ensuring the measures that have been discussed in the findings section, the spread of any pandemic can be reduced in the future. However, this study has some limitations, including the lack of proper data findings. This study can be improved in the future by incorporating a larger dataset and other experimental measurements to analyze the data more accurately.